\documentclass[9pt,conference]{IEEEtran}
\IEEEoverridecommandlockouts
\usepackage{makecell}
\usepackage{cite}
\usepackage{amsmath,amssymb,amsfonts}
\usepackage{float}
\usepackage{booktabs}
\usepackage{multirow}
\usepackage{algorithmic}
\usepackage{graphicx}
\usepackage{textcomp}
\usepackage{xcolor}
\def\BibTeX{{\rm B\kern-.05em{\sc i\kern-.025em b}\kern-.08em
    T\kern-.1667em\lower.7ex\hbox{E}\kern-.125emX}}
\begin{document}

\title{SEF-PNet: Speaker Encoder-Free Personalized Speech Enhancement with Local and Global Contexts Aggregation\\
\thanks{$^{\star}$Yanhua Long is the corresponding author. The work is supported by the National Natural Science Foundation of China (Grant No.62071302).}}

\author{\IEEEauthorblockN{Ziling Huang$^{1}$, Haixin Guan$^{2}$, Haoran Wei$^{3}$, Yanhua Long$^{1\star}$}\\
\IEEEauthorblockA{\textit{$^1$Shanghai Normal University, Shanghai, China},
\textit{$^2$Unisound AI Technology Co., Ltd., Beijing, China}\\
\textit{$^3$Samsung Research America, Mountain View, USA}}}
%\textit{$^3$Department of ECE, University of Texas at Dallas, Richardson, TX 75080, USA}}
% haixinguan@unisound.com, yanhua@shnu.edu.cn

\maketitle
%\vspace{12mm}
\begin{abstract}
Personalized speech enhancement (PSE) methods typically rely on pre-trained speaker verification models or self-designed speaker encoders to extract target speaker clues, guiding the PSE model in isolating the desired speech. However, these approaches suffer from significant model complexity and often underutilize enrollment speaker information, limiting the potential performance of the PSE model. To address these limitations, we propose a novel Speaker Encoder-Free PSE network, termed SEF-PNet, which fully exploits the information present in both the enrollment speech and noisy mixtures. SEF-PNet incorporates two key innovations: Interactive Speaker Adaptation (ISA) and Local-Global Context Aggregation (LCA). ISA dynamically modulates the interactions between enrollment and noisy signals to enhance the speaker adaptation, while LCA employs advanced channel attention within the PSE encoder to effectively integrate local and global contextual information, thus improving feature learning. Experiments on the Libri2Mix dataset demonstrate that  SEF-PNet significantly outperforms baseline models, achieving state-of-the-art PSE performance. Our source code is available at https://github.com/isHuangZiling/SEF-PNet. 
\end{abstract}

\begin{IEEEkeywords}
Personalized Speech Enhancement, Speaker Encoder-free, LCA, SEF-PNet.
\end{IEEEkeywords}

\begin{figure*}[ht]
\centering
\setlength{\abovecaptionskip}{0cm}
\includegraphics[width=0.86\textwidth]{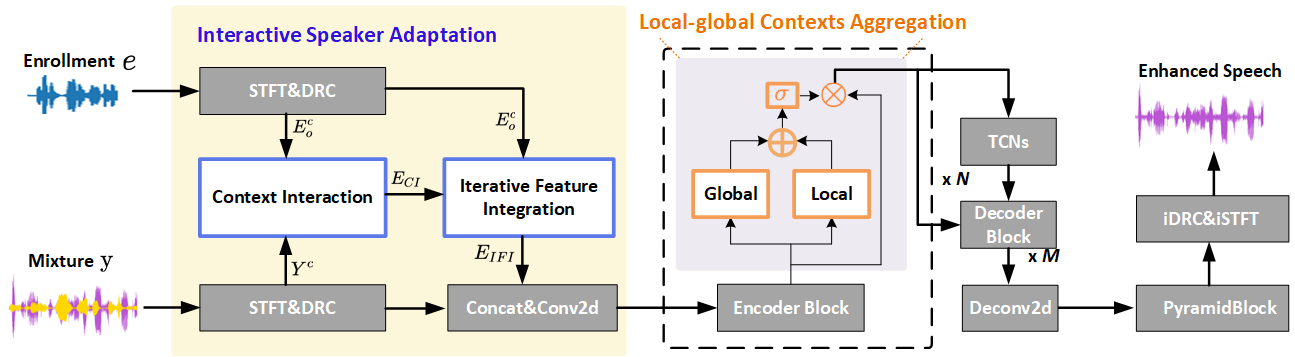}
\caption{Overview of the proposed SEF-PNet model. All colored blocks highlight our key contributions over the original sDPCCN.}
\label{fig:isalca}
%\vspace{-0.3cm}
\end{figure*}

\begin{figure}[!htbp]
\centering
\setlength{\abovecaptionskip}{0cm}
\includegraphics[width=0.42\textwidth]{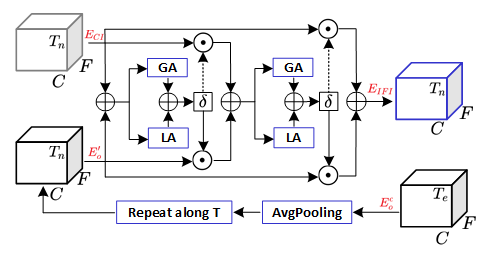}
%\vspace{-0.3cm}
\caption{Structure of the Iterative Feature Integration (IFI) included in the ISA module. GA and LA 
represent Global Attention and Local Attention, respectively, as designed in the LCA module in Fig.\ref{fig:lca}.}
\label{fig:isa}
%\vspace{-0.5cm}
%\setlength{\abovecaptionskip}{-0.5cm}
\end{figure}

\section{Introduction}
Personalized Speech Enhancement (PSE), also known as Target Speaker 
Extraction, aims to extract and enhance the speech of a target speaker in complex multi-talker noisy environments. Research interest in PSE has grown significantly, driven by the specially designed PSE tracks in the Deep Noise Suppression (DNS) Challenges \cite{dubey2024icassp,reddy2021icassp} over the past two years, and its broad applications in real-time communication, speaker diarization, etc. However, state-of-the-art PSE methods still rely on pre-trained speaker verification models or additional self-designed speaker encoders to generate target speaker cues from enrollment speech \cite{tea-pse,tea-pse2.0,tea-pse3.0,x-sepformer,pbsrnn,hierarchical}, which are crucial for guiding the PSE model in extracting the desired speech. These methods not only increase model parameters but also add complexity to PSE inference, limiting their deployment in industrial scenarios. Therefore, developing PSE methods that are free from speaker encoders or embeddings is both necessary and essential.

In the literature, methods for extracting target speaker cues from enrollment speech to guide PSE can be divided into three main categories: 
1) Pre-trained Speaker Verification Models: These methods often use pre-trained models like ECAPA-TDNN\cite{ECAPA-TDNN} or ResNet34\cite{resnet} to extract target speaker embeddings from enrollment speech, which are then transformed and fused with noisy mixture representations in PSE;
2) Self-designed Speaker Encoders: These approaches utilize simple, task-specific speaker encoders to extract and fuse representations from enrollment speech \cite{td-speakerbeam,td-speakerbeaminn,spex,spex+,mc-spex,sdpccn}; 
3) Hierarchical Speaker Fusion: This method combines embeddings from both pre-trained models and self-designed encoders to guide the PSE network hierarchically in extracting the desired speech \cite{hierarchical}. 
Although these methods have achieved great success in many PSE tasks, they suffer from significant model complexity, including large model size and slow inference speed. Moreover, embeddings from pre-trained models may not be optimal for PSE guidance, while self-designed speaker encoders often lack generalization ability to other PSE architectures. These limitations significantly hinder the practical application of these methods in real-world industry.

Therefore, recent research has begun to explore speaker embedding/ encoder-free PSE methods. For example, authors in \cite{sef-net} proposed a fully weight-sharing encoder to capture information from both target speaker enrollment speech and noisy mixtures. However, this approach is model specifically tailored and face scalability issues. Similarly, the work in \cite{yangxue} introduced direct interactions between the time-frequency representations of enrollment and noisy speech, which, while achieving good PSE performance, may still underutilize the target speaker information in the enrollment speech, making it less than optimal.In addition, current PSE models tend to underestimate the significance of channel attention. Although some speech enhancement or separation models\cite{wang2023efficient1,wang2023efficient2,yue2022reference} 
incorporate the squeeze-and-excitation modules (SEblock)\cite{seblock} 
or convolutional block attention module (CBAM)\cite{cbam} to enhance 
feature learning, these approaches primarily focus on global 
information and tend to neglect the local features, thus limiting the system performance.

Motivated by these observations, in this study, we propose a novel PSE network, SEF-PNet, which is designed to effectively capture and learn the information from both enrollment speech and input noisy mixtures. Three main contributions are as follows: 1) \textbf{Speaker Encoder/Embedding Free}: SEF-PNet operates without the need of pre-trained models or self-designed speaker encoders, greatly simplifying the model architecture; 2) \textbf{Interactive 
Speaker Adaptation (ISA)}: ISA enables target speaker adaptation within the PSE encoder through a dynamic interactive manner, ensuring comprehensive information exploration via context interaction and iterative feature integration between the enrollment speech and noisy mixtures; 3) \textbf{Local and Global Context Aggregation (LCA)}: LCA improves  high-level feature learning by integrating local and global contexts across channels within intermediate encoder representations. Our experiments are performed on three differrent Libri2Mix PSE tasks\cite{libri2mix}, and results show that the proposed SEF-PNet consistently outperforms strong baselines and achieves competitive performance compared to other state-of-the-art PSE models.

\section{Proposed Methods}
In this section, we present the details of our proposed speaker encoder/embedding-free PSE approach, including the SEF-PNet architecture, the principle of the proposed interactive speaker adaptation (ISA) and the local-global contexts aggregation (LCA). 

\subsection{Model Architecture}
\label{ssec:SEF-PNet}

The structure of SEF-PNet, as illustrated in Fig.\ref{fig:isalca}, builds upon our previous work, sDPCCN\cite{sdpccn}, which has already demonstrated competitive performance in end-to-end TSE tasks. Using sDPCCN as the backbone, we implemented several key modifications to achieve a speaker encoder-free design. 
First, we removed the original speaker encoder module, simplifying the architecture and focusing on the direct utilization of enrollment speech, as shown in the ISA module (left part of Fig.\ref{fig:isalca}). Second, we integrated the LCA module into the original Encoder Block to exploit both local and global contextual information in the high-level acoustic representations. 
In addition, we reduced the number of Encoder Blocks from 7 to 6 and shortened the frame window length from 64ms to 32ms. These changes preserved the model's performance while reducing its size by approximately 0.7M parameters. The others remain consistent with those in the original sDPCCN. Finally, as shown in Fig.\ref{fig:isalca}, in SEF-PNet, the entire encoder now consists of an ISA adaptation module followed by 6 Encoder Blocks, each enhanced with the LCA module, significantly boosting PSE performance while maintaining much lower model complexity.

\subsection{Interactive Speaker Adaptation (ISA)}

Our inspiration for the ISA comes from \cite{yangxue}, which explores direct context interaction (CI) between Short-Time Fourier Transform (STFT) features of the enrollment speech and the input noisy mixture, using the resulting noisy-adapted enrollment STFT features as the PSE guidance. However, this direct speaker adaptation is heavily tailored to the input mixture, which may underutilize important target speaker clues present in the original enrollment speech. Therefore, as shown in the left part of Fig.\ref{fig:isalca}, we propose an interactive speaker adaptation (ISA) in this study. In the ISA, we enhance the STFT-based CI from \cite{yangxue} by introducing an additional Iterative Feature Integration (IFI) module, which enables soft adaptive weighting to balance the information derived from original enrollment STFT and the noisy-adapted version. This improvement provides better target speaker guidance, enhancing SEF-PNet's PSE performance.  

\begin{figure*}[t]
\centering
\setlength{\abovecaptionskip}{0cm}
\includegraphics[width=0.8\textwidth]{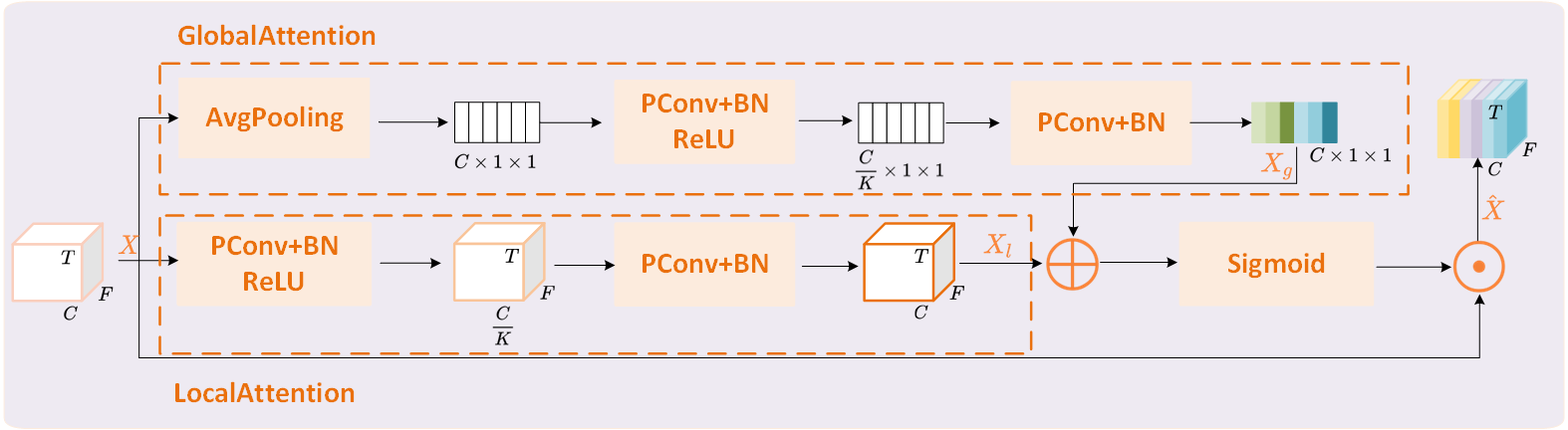}
%\vspace{-0.3cm}
\caption{Structure of Local and Global Contexts Aggregation (LCA) module.}
\label{fig:lca}
%\vspace{-0.4cm}
%\setlength{\abovecaptionskip}{-0.5cm}
\end{figure*}

The CI module in ISA achieves a fully speaker encoder-free process through simple matrix multiplication and its implementation can be formulated as follows: Given the enrollment speech $e$ and  noisy mixture $\mathrm{y}$, they are first transformed into their T-F representations $\mathbf{E}$ and $\mathbf{Y}$ by the STFT, respectively, after applying dynamic range compression (DRC)\cite{drc}, the final compressed STFTs are defined as,
\begin{equation}
\mathbf{\hat{Y}}^c = |\mathbf{Y}|^\beta e^{j\theta_Y},
\end{equation}
\begin{equation}
\mathbf{\hat{E}}_{o}^c = |\mathbf{E}|^\beta e^{j\theta_E},
\end{equation}
where the DRC compressed factor $\beta = 0.5$, $\theta_{E}$, $\theta_{Y}$ 
are the phase of $\mathbf{E}$ and $\mathbf{Y}$, respectively. Then 
the real and imaginary parts of $\mathbf{\hat{Y}}^{c}$ and 
$\mathbf{\hat{E}}_{o}^{c}$ are concatenated to form 
the $\mathbf{Y}^{c}$ and $\mathbf{E}_{o}^{c}$ as CI inputs,
\begin{equation}
\mathbf{Y}^{c} = \left(\mathbf{\hat{Y}}_{R}^{c}, \mathbf{\hat{Y}}_{I}^{c} \right),  \ \ \  \mathbf{E}_{o}^{c} = \left(\mathbf{\hat{E}}_{oR}^{c}, \mathbf{\hat{E}}_{oI}^{c} \right)
\end{equation}
Next, the context interaction is performed by calculating the noisy-adapted target speaker guidance $\mathbf{E}_{CI}$ using the following equation,
\begin{equation}
\mathbf{E}_{CI} = \texttt{softmax}\left (\mathbf{Y}^{c} \times (\mathbf{E}_{o}^{c})^{\mathrm{T} } \right) \times \mathbf{E}_{o}^{c}
\end{equation}
where $\mathrm{T}$ is the matrix transpose.

Since the output of CI module, $\mathbf{E}_{CI}$, represents the noisy-adapted representation of target speaker clues, may lead to underutilization of the enrollment speech. Therefore, in the proposed IFI module of ISA, we take both $\mathbf{E}_{CI}$ and the enrollment $\mathbf{E}_{o}^c$  as inputs. As illustrated in Fig.\ref{fig:isa}, to enhance the global capture of target speaker cues, we first transform $\mathbf{E}_{o}^c$ into $\mathbf{E}'_{o}$. This is done by first applying average pooling along the time frame dimension, followed by repeating the resulting vector along the time dimension to match the shape of $\mathbf{E_c}$. Next, a mask probability matrix $\mathbf{P}$ is caculated as follows, 
\begin{equation}
\mathbf{P}  = \sigma(\texttt{GA}(\mathbf{E}_{CI} + \mathbf{E}'_{o}) + \texttt{LA}(\mathbf{E}_{CI} + \mathbf{E}'_{o})) 
\end{equation}
where $\sigma(\cdot)$ is the sigmoid activation function, 
$\texttt{GA}$ and $\texttt{LA}$ represent the operations of 
Global Attention and Local Attention, respectively, as designed in the LCA module in Fig.\ref{fig:lca}. This mask $\mathbf{P}$ is then 
used to adaptively weight $\mathbf{E}_{CI}$ and $\mathbf{E}'_{o}$, producing the output of the first round of iterative feature integration, $\mathbf{E}_{CI}^{o}$:  
\begin{equation}
\mathbf{E}_{CI}^{o}  = \mathbf{P} \odot \mathbf{E}_{CI} + (\mathbf{1} - \mathbf{P}) \odot  \mathbf{E}'_{o}
\end{equation}
where $\odot$ denotes element-wise multiplication. 
The second iteration of feature integration then takes $\mathbf{E}_{CI}^{o}$ as input, yielding the final output of the IFI module, $\mathbf{E}_{IFI}$: 
\begin{equation}
\begin{aligned}
\mathbf{E}_{IFI} & = \sigma(\texttt{GA}(\mathbf{E}_{CI}^{o} + \texttt{LA}(\mathbf{E}_{CI}^{o}) \odot \mathbf{E}_{CI} \\ 
& + (\mathbf{1} - \sigma(\texttt{GA}(\mathbf{E}_{CI}^{o} + \texttt{LA}(\mathbf{E}_{CI}^{o}))) \odot \mathbf{E'_o}
\end{aligned}
\end{equation}
 
By combing context interaction with the  interactive  feature integration in ISA, $\mathbf{E}_{IFI}$ effectively captures both the input noisy acoustic characteristics and fully exploits the target speaker cues from the enrollment speech, thereby providing comprehensive and improved target speaker guidance for Personalized Speech Enhancement.

\subsection{Local and Global Contexts Aggregation (LCA)}

Fig.\ref{fig:lca} illustrates the architecture of  LCA module, designed to enhance the feature representation of the encoder in our SEF-PNet by capturing both local and global contextual information.  
Given an input feature map $\mathbf{X}$ with dimensions $C \times T \times F$, the LCA processes it using two parallel attention paths: 
Global Attention (\texttt{GA}) and Local Attention (\texttt{LA}). 
The operations of \texttt{GA} is shown in the upper part of Fig.\ref{fig:lca}. 
First, it utilizes average pooling (\texttt{AvgPooling}) across channels to produce a feature map of size $C\times 1 \times 1$. The pooled feature is then passed through a point-wise convolution (\texttt{PConv}) layer, followed by batch normalization (\texttt{BN}) and ReLU activation. This layer reduces the dimensionality to $C/K$ channels. The output is subsequently processed by another \texttt{PConv+BN} layer, resulting in the global context vector $\mathbf{X}_{g}$. Simultaneously, in the \texttt{LA} module (lower part of Fig.\ref{fig:lca}), the input $\mathbf{X}$ undergoes a similar transformation through a \texttt{PConv+BN+ReLU} operation, followed by a \texttt{PConv+BN} layer to preserve local temporal patterns, producing the local context feature map $\mathbf{X}_{l}$.  

Next, LCA aggregates the contextual information through a fusion and attention (FA) mechanism, followed by feature recalibration. In the FA, $\mathbf{X}_{l}$ is combined with $\mathbf{X}_{g}$ through element-wise addition. This fused representation is then passed through a Sigmoid activation to generate an attention mask, which serves as the core of the FA. Finally, in the feature recalibration step, the input $\mathbf{X}$ is element-wise multipliedby the generated attention mask, producing the recalibrated output $\mathbf{\hat{X}}$. This output integrates both local and global contextual information, enhancing the feature map of each encoder in SEF-PNet and facilitating improved PSE performance. 

\section{Experiments and Results}

\subsection{Datasets}\label{datasets}

In this work, all our experiments are conducted on three public PSE conditions from the Libri2Mix dataset \cite{libri2mix}: 
1) 1-speaker+noise (\texttt{mix\_single}): the mixture signal with only a single target speaker and background noise;  
2) 2-speaker (\texttt{mix\_clean}): the mixture with a target speaker and one interfering speaker without any additional noise. 
3) 2-speaker+noise (\texttt{mix\_both}): the mixture with a target speaker, one interfering speaker and background noise. The \texttt{mix\_*} names are original PSE conditions in Libri2Mix, For clarity, we use the terms `1-speaker+noise', `2-speaker' and `2-speaker+noise' as representative labels instead of the original ones. In each condition, the training set includes 13,900 utterances from 251 speakers, while both the development and test sets contain 3,000 utterances from 40 speakers each, with all mixtures simulated in the `minimum' mode. All mixtures are resampled to 8 kHz. Note that only the first speaker is taken as the target speaker during all training mixture data simulation (the same data preparation as our baseline sDPCCN in \cite{sdpccn} ), unless otherwise specified.

\subsection{Models}\label{models}

The baseline model, sDPCCN \cite{sdpccn}, serves as the backbone of our proposed SEF-PNet. Its primary structure includes a speaker encoder, 7 Encoder Blocks, and 7 corresponding Decoder Blocks, which are connected by a TCNs module, followed by a PyramidBlock and a Deconv2d layer. Each Encoder Block consists of a DenseBlock, 4 DenseEncoders, and 3 Conv2dBlocks. The TCNs module features two layers, each containing 10 TCN blocks. Each Decoder Block is composed of 3 Deconv2dBlocks, 4 DenseDecoders, and a DenseBlock. The PyramidBlock is constructed using average pooling and four parallel conv2d + upsample operations, which are concatenated to form the final output. Further details of the sDPCCN model can be found in our previous work \cite{sdpccn}. The structural and detailed differences between our proposed SEF-PNet and the sDPCCN model are discussed in Section \ref{ssec:SEF-PNet}. In the ISA and LCA modules, all convolutional kernels and strides are set to 1, with channel upsampling or downsampling factors $K=1/32$ and $K=4$, respectively.  

\subsection{Configurations}\label{exps}

We employ the Adam\cite{adam} optimizer with an initial learning rate of 0.0005. The learning rate is adjusted by multiplying it by 0.98 every two epochs for the first 100 epochs and by 0.9 in the last 20 epochs. Gradient clipping\cite{zhang2019gradient} is applied to limit the maximum L2-norm to 1. The training procedure lasts for up to 120 epochs. The mdoel training objective is to maximize the scale-invariant signal-to-distortion ratio (SI-SDR) \cite{sisdr} between the enhanced and the ground-truth signals of the target speaker.Three metrics: SI-SDR(dB), PESQ\cite{pesq} and STOI (\%)\cite{stoi} are used for evaluating our PSE system performance.

\subsection{Ablation Results on Libri2Mix 2-speaker 
Condition}\label{results}

Table \ref{tab:2spkpse} presents the results of ablation study on the Libri2Mix 2-speaker PSE condition, highlighting the impact of each key component proposed in the SEF-PNet model. Comparing S1 to S0, we observe that replacing the speaker encoder in the sDPCCN  with the CI module greatly enhances performance across all metrics, even with the model size reduced by 0.7M. Further improvements are observed in S2, where the integration of the IFI module into S1 leads to notable performance gains, suggesting that the combination of IFI and CI within the ISA module effectively strengthens target speaker guidance.In addition, in S3, the introduction of the LCA module demonstrates its effectiveness in aggregating local and global contexts within the encoder. As a result, the proposed SEF-PNet substantially outperforms the baseline S0, with SI-SDR improving from 11.62 to 13.00, PESQ from 2.76 to 3.05, and STOI from 87.19\% to 89.71\%.

\begin{table}[H]
\vspace{-1em}
\renewcommand\arraystretch{1.0}
\caption{Results on Libri2Mix 2-speaker PSE condition. }
  \label{tab:2spkpse}
  \centering
  \scalebox{1.0}{
	\begin{tabular}{l|l|c|c|c|c|c}
		\toprule
		\textbf{ID} & \textbf{Methods} & \textbf{SI-SDR} & \textbf{PESQ} & \textbf{STOI} & \textbf{Params} & \textbf{MACs} \\
		\midrule
		- & Mixture & -0.03 & 1.60 & 71.38 & - & - \\
		S0 & \textbf{sDPCCN} & \textbf{11.62} & \textbf{2.76} & \textbf{87.19} & \textbf{6.63M} & \textbf{8.49G} \\
   % \hline
		S1 & DPCCN+CI & 12.11 & 2.87 & 88.47 & 5.89M & 8.37G \\
		S2 & + IFI & 12.81 & 2.93 & 89.68 & 5.89M & 8.43G \\
		%S3 & + LCA & 12.89 & 3.00 & 0.90 & 6.07M \\
		\multirow{2}{*}{S3} & ++ LCA & \multirow{2}{*}{\textbf{13.00}} & \multirow{2}{*}{\textbf{3.05}} & \multirow{2}{*}{\textbf{89.71}} & \multirow{2}{*}{\textbf{6.08M}} & \multirow{2}{*}{\textbf{8.50G}} \\
        & \textbf{(SEF-PNet)} &  &  &  &  &  \\ 
		\bottomrule
	\end{tabular}}
\end{table}

Table \ref{tab:lcarst} compares the proposed LCA with the commonly used 
SEblock \cite{seblock} and CBAM \cite{cbam} within the S1 configuration to 
examine their effectiveness of different context aggregation methods on the 
Libri2Mix 2-speaker condition. The results show that the SEblock performs better than CBAM, but our LCA outperforms both. This superiority may be attributed to the fact that SEblock and CBAM primarily focus on capturing global contexts, whereas the LCA is designed to aggregate information at both local and global scales, enabling it to achieve the best results.

\begin{table}[H]
\vspace{-1em}
\renewcommand\arraystretch{1.0}
\caption{Comparison with different context aggregation methods. }
  \label{tab:lcarst}
  \centering
  \scalebox{1.0}{
	\begin{tabular}{l|c|c|c}
		\toprule
		\textbf{Methods} & \textbf{SI-SDR} & \textbf{PESQ} & \textbf{STOI}  \\
		\midrule
		S1  & 12.11 & 2.87 & 88.47 \\
		+ LCA & 12.88 & 3.00 & 89.70  \\
		+ SEblock & 12.63 & 2.92 & 89.51  \\
		+ CBAM & 12.50 & 2.91 & 89.08 \\
		\bottomrule
	\end{tabular}}
\end{table}

\subsection{Condition-wise Results on Three PSE Conditions}

Table \ref{tab:rst3pse} shows the performance of SEF-PNet compared to the baseline sDPCCN on Libri2Mix three different PSE conditions. It's clear to observe that except for the similar performance on `1-speaker+noise' condition, the proposed SEF-PNet consistently outperforms sDPCCN across two other conditions with speaker interference, demonstrating its strong effectiveness and generalization ability in enhancing diversity PSE performance. Moreover, we see that both sDPCCN baseline and proposed SEF-PNet perform well on `1-speaker+noise' and `2-speaker' PSE tasks, and relatively poor on `2-speaker+noise' condition. This is due to the fact that `2-speaker+noise' condition is more complicated than other two conditions because the target speech is deteriorated by both interfering speaker and noise. 

\begin{table}[H]
\vspace{-0.5em}
\renewcommand\arraystretch{1.0}
\caption{Condition-wise results on three Libri2Mix PSE tasks.}
\label{tab:rst3pse}
\centering
\scalebox{1.0}{
\begin{tabular}{l|l|c|c|c} 
\toprule
\textbf{Condition} & \textbf{Method} & \textbf{SI-SDR} & \textbf{PESQ} & \textbf{STOI} \\
%& & \textbf{SISDRi} & \textbf{PESQi} & \textbf{STOIi} \\
\midrule
\multirow{3}{*}{1-speaker+noise} & Mixture  & 3.27 & 1.75 & 79.51 \\
                          & sDPCCN   & 14.49 & 3.04 & 92.47 \\
                          & SEF-PNet & 14.50 & 3.05 & 92.47  \\
\midrule
\multirow{3}{*}{2-speaker} & Mixture  & -0.03 & 1.60 & 71.38 \\ 
                          & sDPCCN   & 11.62 & 2.76 & 87.19 \\
                          & SEF-PNet & 13.00 & 3.05 & 89.71 \\
\midrule
\multirow{3}{*}{2-speaker+noise} & Mixture  & -2.03 & 1.43 & 64.65 \\ 
                          & sDPCCN   & 6.93  & 2.12 & 79.32 \\
                          & SEF-PNet & 7.54   & 2.14  & 80.58 \\
\bottomrule
\end{tabular}}
\end{table}

\subsection{Comparison with Existing Methods}

Table \ref{tab:cmpother} demonstrates the performance comparison between 
our proposed SEF-PNet and two other competitive state-of-the-art (SOTA) PSE methods on the 2-speaker condition of Libri2Mix. To ensure a fair comparison with SpEx+\cite{spex+} and MC-SpEx\cite{mc-spex}, we trained the SEF-PNet model using the same amount of simulated training data. During the training data simulation, each speaker in the mixture was alternately treated as the target speaker, doubling the size of the simulated training data compared to what was used in  Table \ref{tab:2spkpse} to Table \ref{tab:rst3pse}. From the results in Table \ref{tab:cmpother}, it is clear that our proposed SEF-PNet achieves SOTA performance across all evaluation metrics, while also having a significantly smaller model size, highlighting its superiority over the other two competitive methods.

\begin{table}[H]
\vspace{-1em}
\renewcommand\arraystretch{1.0}
\caption{Performance comparison with various existing PSE methods on Libri2Mix 2-speaker condition (training data size $\times 2$).}
  \label{tab:cmpother}
  \centering
  \scalebox{1.0}{
	\begin{tabular}{l|c|c|c|c|c}
		\toprule
		\textbf{Methods} & \textbf{Params} & \textbf{SI-SDR} & \textbf{PESQ} & \textbf{STOI} & \textbf{MACs} \\
		\midrule
		Mixture & - & 0.00 & 1.60 & 71.29 & - \\
		SpEx+\cite{spex+} & 11.78M & 13.41 & 2.93 & - & 10.73G \\
        MC-SpEx\cite{mc-spex} & 10.77M & 14.61 & 3.19 & - & - \\
		\midrule
		\textbf{SEF-PNet} & \textbf{6.07M} & \textbf{14.61} & \textbf{3.19} & \textbf{91.75} & \textbf{8.50G} \\
		\bottomrule
	\end{tabular}}
\end{table}

\subsection{Visualization}

\begin{figure}[!htbp]
\centering
\setlength{\abovecaptionskip}{0pt}
\includegraphics[width=0.5\textwidth]{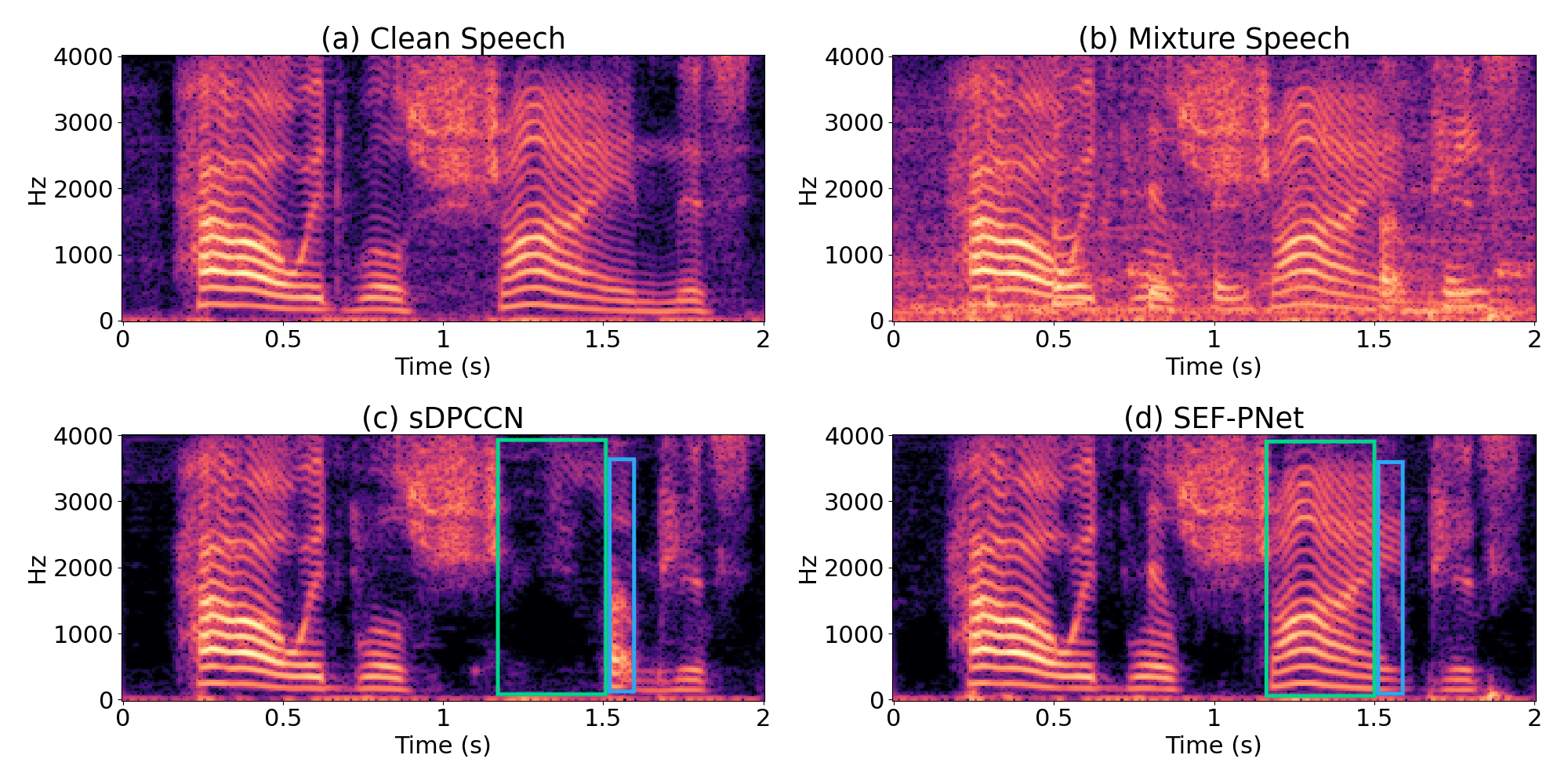}
\caption{Spectrograms of (a) clean speech, (b) mixture speech,
(c) sDPCCN and (d) SEF-PNet.}
\label{fig:visal}
\vspace{-0.4cm}
\end{figure}

Fig.\ref{fig:visal} illustrates the comparison in spectrograms of the same noisy mixture that randomly selected from the test set of the Libri2Mix 2-speaker+noise condition, enhanced by the sDPCCN baseline and our proposed SEF-PNet. It is clear that SEF-PNet achieves better noise suppression and more effective removal of interfering speakers compared to sDPCCN, thereby enhancing the overall PSE performance.

\section{Conclusion}

This paper introduces SEF-PNet, a Speaker Encoder/embedding-Free Personalized Speech Enhancement network that eliminates the need for complex speaker encoders or pre-trained speaker verification models. By leveraging the proposed Interactive Speaker Adaptation (ISA) and Local-Global Context Aggregation (LCA) mechanisms, SEF-PNet not only performs excellent direct interactive target speaker adaptation, but also effectively aggregates both local and global contextual information during feature learning. Experimental results across three Libri2Mix PSE conditions proved that SEF-PNet significantly outperforms our sDPCCN baseline, achieving state-of-the-art performance with fewer parameters. Our future work will focus on generalizing the proposed ISA and LCA techniques to other PSE architectures and scaling them to larger industry PSE tasks.

\end{document}